\documentclass[aps,pre,showpacs,showkeys,preprintnumbers,floatfix,nofootinbib,12pt]{revtex4-1}
\pdfoutput=1
\usepackage{amsmath}
\usepackage{amssymb}
\usepackage{graphicx,color}
\usepackage{multirow}

\begin{document}

\title{Mapping the phase transitions of the ZGB model with inert sites via
nonequilibrium refinement methods}
\author{Henrique A. Fernandes$^{1}$, Roberto da Silva$^{2}$}

\begin{abstract}
In this paper, we revisit the ZGB model and explore the effects of the
presence of inert sites on the catalytic surface. The continuous and
discontinuous phase transitions of the model are studied via time-dependent
Monte Carlo simulations. In our study, we are concerned with building a
refinement procedure, based on a simple concept known as coefficient of
determination $r$, in order to find the possible phase transition points
given by the two parameters of the model: the adsorption rates of carbon
monoxide, $y$, and the density of inert sites, $\rho_{is}$. First, we obtain 
$10^6$ values of $r$ by sweeping the whole set of possible values of the
parameters with an increment $10^{-3}$, i.e., $0 \leq y \leq 1$ and $0 \leq
\rho_{is} \leq 1$ with $\Delta y=\Delta\rho_{is}=10^{-3}$. Then, with the
possible phase transition points in hand, we turn our attention to some
fixed values of $\rho_{is}$ and perform a more detailed refinement
considering larger lattices and increasing the increment $\Delta y$ by one
order of magnitude to estimate the critical points with higher precision.
Finally, we estimate the static critical exponents $\beta$, $\nu_{\parallel}$%
, and $\nu_{\perp}$, as well as the dynamic critical exponents $z$ and $%
\theta$.
\end{abstract}

\maketitle

\affiliation{$^1$Instituto de Ci{\^e}ncias Exatas, Universidade Federal de Goi{\'a}s, Regional Jata{\'i}, BR 364, km 192, 3800 - CEP 75801-615, Jata{\'i}, Goi{\'a}s, Brazil \\
$^2$Instituto de F{\'i}sica, Universidade Federal do Rio Grande do Sul, Av. Bento Gon{\c{c}}alves, 9500 - CEP 91501-970, Porto Alegre, Rio Grande do Sul, Brazil}

\section{Introduction}

\label{sec:introduction}

The production of carbon dioxide from catalytic surfaces is of fundamental
interest from both scientific and technological points of view. In this
context, one can consider the model devised by Ziff, Gulari, and Barshad 
\cite{ziff1986} in 1986 which, despite its simplicity, presents two
different phase transitions: one continuous and another discontinuous. In
addition, on the contrary of the continuous phase transition which seems to
be only a theoretical prediction, the discontinuous one has been
experimentally verified by several authors. The only parameter of the model
is the adsorption rate of carbon monoxide molecules.

An important question that arises when studying such simple models is: How
are the phase transitions affected by the inclusion of other parameters? As
presented below, this question has been considered by several authors in
recent years. For instance, some authors have included the diffusion of the
adsorbed species (carbon monoxide molecules or oxygen atoms) on the
catalystic surface \cite{jensen1990a, Chan2015, buendia2015, Ewans1993,
dasilva2018}, and others have studied the model with the desorption of
molecules from the surface \cite{tome1993, albano1992, brosilow1992,
kaukonen1989, fernandes2018}. In this paper, we include to the original
model the existence of inert sites (or impurities) over the surface (see,
for example, Ref. \cite{hoenicke2014}). So, our intent is to analyse the two
phase transitions of the original model when the adsorption rate and the
density of inert sites vary. This study is focused on the use of a
refinement method proposed for nonequilibrium Monte Carlo simulations in the
context of short-time dynamics rather than the steady state Monte Carlo
simulations. In Ref. \cite{dasilva2018}, we have shown that both continuous
and discontinuous phase transitions are preserved when the diffusion of the
adsorbed species are allowed. On the other hand, we have shown in Ref. \cite%
{fernandes2018} that the introduction of the desorption of carbon monoxide
molecules from the lattice preserves the continuous phase transition and
destroys the discontinuous one even for small values of desorption rates. In
that same work, we have also shown that before the dissapearance of the
discontinuous phase transition, there exists a sequence of ``pseudocritical
points'' forming two lines that end in a single point. As predicted by Tom%
\'{e} and Dickman \cite{tome1993}, we have confirmed through the dynamic
critical exponent $\theta $ that this is an Ising-like critical point.

In this paper, we are interested in showing how the insertion of the density
of inert sites can affect the continuous and discontinuous phase transitions
of the ZGB model. First, we analyze the behavior of the densities of CO and
CO$_{2}$ molecules, of O atoms, and of vacant sites according to the two
parameters of the model: the density of inert sites and the adsorption rate
of CO molecules on the surface. To reach this goal, we consider a
non-traditional way to explore such transitions. This approach, which was
proposed in 2012 for generalized systems \cite{silva2012}, considers a
refinement method based on a simple statistical concept known as coefficient
of determination. It has been succesfully applied in models without defined
Hamiltonian \cite{silva2015, fernandes2016, dasilva2018, fernandes2018}, in
models with defined Hamiltonian and with short-range interactions \cite%
{silva2012,silva2013a,silva2013b,silva2014,fernandes2017}, and recently in
long-range systems \cite{Longrange}. With this method, we are able
to obtain diagrams with the possible regions of phase transitions for the
whole set of possible of values of adsorption rates and densities of inert
sites of the model. After identifying the possible phase transitions points
of the model through the coefficient of determination, we focused our
attention to some determined points and refine our measurements in order to
verify with good precision the influence of inert sites on the ZGB model.
So, for the first time, the refinement process is carried out on two levels:
at the first level, we consider a smaller lattice size and sweep all
possible values of adsorption rates and densities of inert sites in order to
obtain an overview of the phase transitions of the model; on the second
level, we focus our attention only on the regions close to the phase
transition points obtained previously in order to calculate the coefficient
of determination for larger system sizes besides improving the precision of
the measurements in one magnitude order (fine scale). Therefore, the process
is even more precise than other previous explorations of the method.
Finally, we calculate several critical exponents for some points in order to
check the universality class of the model.

In the next section, we present the model to be studied and in Sec. \ref%
{sec:method} we briefly show the nonequilibrium Monte Carlo method as well
as the refinement procedure proposed in 2012 \cite{silva2012} and employed
here to estimate the phase transition points of the model according to the
CO adsorption rate and the density of inert sites. In Sec. \ref{sec:results}%
, we present our main results and our conclusions are presented in Sec. \ref%
{sec:conclusions}.

\section{the model}

\label{sec:model}

In 1986, Ziff, Gulari, and Barshad \cite{ziff1986} devised a simple, but at
the same time very interesting, model addressing the production of carbon
dioxide molecules (CO$_{2}$) from catalytic surfaces. In their model, also
known as ZGB model, the catalytic surface is represented by a regular square
lattice whose sites can be filled with oxygen atoms (O), carbon monoxide
molecules (CO), or be vacant (V). Both CO and O$_{2}$ molecules in the gas ($%
g$) phase are able to impinge the surface with a rate $y$ and $1-y$,
respectively. Therefore, $y$ is the only parameter that controls the
kinetics of the model. If the chosen molecule in the gas phase is CO, it is
adsorbed ($a$) on the surface if a site, chosen at random, is empty.
Otherwise, if the O$_{2}$ molecule is chosen, it dissociates into two O
atoms and both are adsorbed on the surface only if the two nearest-neighbor
sites, also randomly chosen, are vacant. If any of the adsorption sites is
occupied, the adsorption processes do not occur and the molecules return to
the gas phase. The catalytic reaction, which produces CO$_{2}$ molecules ($%
\text{CO}+\text{O}\longrightarrow \text{CO}_{2}$) in the gas phase ($g$),
occurs whenever the O atoms and CO molecules adsorbed on the surface are
nearest neighbors. This set of reactions follows the Langmuir-Hinshelwood
mechanism \cite{ziff1986, evans1991} and can be represented by the following
reaction equations: 
\begin{align}
\text{CO}(g)+\text{V}& \longrightarrow \text{CO}(a) \\
\text{O}_{2}(g)+2\text{V}& \longrightarrow 2\text{O}(a) \\
\text{CO}(a)+\text{O}(a)& \longrightarrow \text{CO}_{2}(g)+2\text{V}
\label{eq:reaction}
\end{align}
As we stated above, despite its simplicity, this model is very interesting
since it possesses two different irreversible phase transitions (IPT)
separating two absorbing states from an active phase where there is the
production of CO$_{2}$ molecules. One of these transitions is continuous and
occurs at $y_{1}\simeq 0.3874$ \cite{voigt1997}. It separates the absorbing
state ($0\leq y<y_{1}$), where the whole surface is poisoned with $O$ atoms,
from the active phase. The other transition occurs at $y_{2}\simeq 0.5256$ 
\cite{ziff1992}, and is discontinuous. At this point, the production of CO$%
_{2}$ molecules ceases and the system reaches the other absorbing state ($%
y_{2}<y\leq 1$) where all sites on the lattice are filled with CO molecules.
In the active phase ($y_{1}<y<y_{2}$), both CO molecules, O atoms, and
vacant sites coexist on the catalytic surface with sustainable production of
CO$_{2}$ molecules.

These properties alone are sufficient to become the ZGB model a prototype in
numerical studies of reaction processes on catalytic surfaces. In addition,
some experimental works on platinum confirm the existence of discontinuous
IPT in the catalytic oxidation of CO molecules \cite{golchet1978,
matsushima1979, ehsasi1989, christmann1991, block1993, imbhil1995}, which
also justify the number of works related to the model presented in
literature since its discovery. Those studies have been performed through
several techniques such as series analysis, mean-field theory and
simulations etc. \cite{marro1999}, along with several improvements proposed
in order to make the model more realistic.

In this work, we consider the ZGB model modified to include inert sites
randomly distributed on the surface, which in turn, can be thought of as
impurities presented on the lattice. The inert sites are chosen at the
beginning of the time evolution of the system and remain fixed for the
considered sample. The adsorption and reaction processes, which follow Eqs.
(1)-(\ref{eq:reaction}) presented above for the original ZGB model, start
only after the inert sites are distributed over the lattice. Therefore, we
set in our simulations the density of inert sites $\rho _{is}$ as another
control parameter of the model (along with the CO adsorption rate $y$). The
study was carried out through numerical simulations for $0\leq y\leq 1$ and $%
0\leq \rho _{is}\leq 1$ with $\Delta y=\Delta \rho _{is}=10^{-3}$ totaling $%
10^{6}$ independent simulations for the points ($y,\rho_{is}$). Our intent
is to look into the whole phase diagram of the model to observe the
influence of inert sites on the phase transitions, critical exponents, and
universality class of the model.

\section{Monte Carlo simulations and the refinement method}

\label{sec:method}

In order to reach our goal, we have considered the well-established
time-dependent Monte Carlo (MC) technique (see, for example, Ref. \cite%
{shorttime}) used in the study of critical phenomena of systems with and
without defined Hamiltonians along with a refinement procedure known as
coefficient of determination. Such a procedure is derived from the
well-known short-time dynamics proposed in 1989 by Janssen \textit{et al.} 
\cite{janssen1989} through renormalization group techniques, and by Huse 
\cite{huse1989} via numerical simulations. They showed that there is
universality and scaling behavior even at the beginning of the time
evolution of dynamical systems at criticality. For systems with absorbing
states, this finding can be translated into the following general scaling
relation \cite{hinrichsen2000,Rdasilva-contact2004}: 
\begin{equation}
\left\langle \rho (t)\right\rangle \sim t^{-\beta /\nu
_{\parallel}}f((y-y_{c})t^{1/\nu _{\parallel }},t^{d/z}L^{-d},\rho
_{0}t^{\beta /\nu_{\parallel }+\theta }),  \label{eq:fss}
\end{equation}
where $\rho (t)=\rho _{\text{V}}(t)$, the density of vacant sites, is the
order parameter of the model which is defined as 
\begin{equation}
\rho (t)=\frac{1}{L^{d}}\sum_{i=1}^{L^{d}}s_{i},  \label{eq:op}
\end{equation}%
where $s_{i}=1\ (0)$ when the sites $i$ are vacant (filled with O atoms or
with CO molecules). In Eq. (\ref{eq:fss}), $\left\langle \cdots\right\rangle 
$ means the average on different evolutions of the system, $d$ is the
dimension of the system, $L$ is the linear size of a regular square lattice,
and $t$ is the time. The exponents $\beta$, $\nu_{\parallel}$, and $%
\nu_{\perp }$ are static critical exponents, $z=\nu_{\parallel }/\nu_{\perp}$
and $\theta =\frac{d}{z}-\frac{2\beta}{\nu_{\parallel}}$ are dynamic ones,
and $y_{c}$ is the critical point, i.e., the critical adsorption rate of CO
molecules.

For this technique, the choice of initial conditions of the system at
criticality is crucial. Here, in order to estimate the critical exponents,
we are able to take into account two different initial conditions, as shown
in Refs. \cite{hinrichsen2000, fernandes2016}. In the first one, all
available sites of the lattice are initially vacant, and from Eq. (\ref%
{eq:fss}), it is expected that the density of vacant sites decays
algebraically as 
\begin{equation}
\left\langle \rho (t)\right\rangle \sim t^{-\beta /\nu _{\parallel }}.
\label{eq:p1}
\end{equation}
Secondly, when the simulation starts with all sites of the lattice filled
with O atoms, except for a single empty site chosen at random, the Eq. (\ref%
{eq:fss}) leads to 
\begin{equation}
\left\langle \rho (t)\right\rangle \sim \rho_0t^{\frac{d}{z}-2\frac{\beta}{%
\nu_{\parallel}}}=\rho_0t^{\theta }.  \label{eq:p2}
\end{equation}

Hence, in $\log \times \log $ scale, the slopes of the power laws given by
Eqs. (\ref{eq:p1}) and (\ref{eq:p2}) are precisely the exponents $\beta
/\nu_{\parallel}$ and $\theta$, respectively.

The dynamic critical exponent $z$ can be found independently when one mixes
these two initial conditions leading to the following power law behavior 
\cite{Rdasilva-contact2004, silva2002a}: 
\begin{equation}
F_2(t)=\left\langle \rho \right\rangle_{\rho_0=1/L}(t)/\left\langle \rho
\right\rangle_{\rho_0=1}^2(t)\sim t^{d/z}.  \label{eq:f2}
\end{equation}
In addition, the exponent $\nu_{\parallel}$ can also be found independently
when considering the derivative $D(t)=\frac{\partial \ln \left\langle \rho
\right\rangle}{\partial y}\bigg\vert_{y=y_c}$ which yields \cite%
{grassberger1996} 
\begin{equation}
D(t)=t^{\frac{1}{\nu_{\parallel}}}.  \label{eq:dv}
\end{equation}
From these power laws, it is possible to obtain the exponents $z$, $\theta$, 
$\beta$, $\nu_{\parallel}$, and $\nu_{\perp}$ separately, without the
problem of critical slowing down characteristic from steady state
simulations.

These power laws are observed only when the system is at criticality.
Therefore, to take advantage of this technique, we need to know, in
principle, the critical parameters of the model with good precision.
However, we can use this technique to localize and refine the critical
parameters by considering a refinement method proposed in 2012 by da Silva 
\textit{et al.} \cite{silva2012}. This approach, which is based on the
refinement of the coefficient of determination of the order parameter allows
to locate phase transitions of systems in a very simple way.

The coefficient of determination is a very simple concept used in linear
fits, or other fits (for more details, see for example, Ref. \cite%
{trivedi2002}). So, let us briefly explain such a procedure in the context
of short-time Monte Carlo simulations. When we perform least-square linear
fit to a given data set, we obtain a linear predictor $\widehat{y}%
_{t}=a+bx_{t}$. In addition, if we consider the unexplained variation given
by 
\begin{equation*}
\widetilde{\Delta }=\sum_{t=1}^{N}(y_{t}-\widehat{y}_{t})^{2},
\end{equation*}
a perfect fit is achieved when the curve is given by $y_{t}=a+bx_{t}$, and
therefore, $\widetilde{\Delta }=0$.

On the other hand, the explained variation $\Delta $ is given by the
difference between the average $\overline{y}=N^{-1}\sum_{t=1}^{N}y_{t}$, and
the prediction $\widehat{y}_{t}$, i.e., 
\begin{equation}
\Delta =\sum_{t=1}^{N}(\widehat{y}_{t}-\overline{y})^{2}.
\end{equation}

So, it is interesting to consider the total variation, naturally defined as 
\begin{equation}
\Delta_{total}=\sum_{t=1}^{N}(y_{t}-\overline{y})^{2}
\end{equation}

So, we can rewrite this last expression as 
\begin{equation}
\begin{array}{lll}
\Delta _{total} & = & \sum_{t=1}^{N}(y_{t}-\overline{y})^{2} \\ 
&  &  \\ 
& = & \sum_{t=1}^{N}(y_{t}-\widehat{y}_{t}+\widehat{y}_{t}-\overline{y})^{2}
\\ 
&  &  \\ 
& = & \sum_{t=1}^{N}(y_{t}-\widehat{y}_{t})^{2}+\sum_{t=1}^{N}(\widehat{y}%
_{t}-\overline{y})^{2}+\xi%
\end{array}%
\end{equation}
where $\xi =2\sum_{t=1}^{N}(y_{t}-\widehat{y}_{t})(\widehat{y}_{t}-\overline{%
y})$. However we can easily show that $\xi =0$, since

\begin{equation}
\begin{array}{lll}
\sum_{t=1}^{N}(y_{t}-c_{a}-c_{b}x_{t})(c_{a}+c_{b}x_{t}-\overline{y}) & = & 
c_{b}\sum_{t=1}^{N}x_{t}(y_{t}-c_{a}-c_{b}x_{t})+(c_{a}-\overline{y}%
)\sum_{t=1}^{N}x_{t}(y_{t}-c_{a}-c_{b}x_{t}) \\ 
&  &  \\ 
& = & -\frac{c_{b}}{2}\frac{\partial }{\partial c_{b}}%
\sum_{t=1}^{N}(y_{t}-c_{a}-c_{b}x_{t})^{2}-\frac{(c_{a}-\overline{y})}{2}%
\frac{\partial }{\partial c_{a}}\sum_{t=1}^{N}(y_{t}-c_{a}-c_{b}x_{t})%
\end{array}%
\end{equation}
and the last two sums vanish by definition when take the least squares
values $(c_a,c_b)=(a,b)$ \cite{trivedi2002}.

Therefore, the total variation can be simply defined as 
\begin{equation}
\Delta _{total}=\widetilde{\Delta }+\Delta
\end{equation}
and the better the fit, the smaller the $\widetilde{\Delta }$. So, in an
ideal situation $\widetilde{\Delta}=0$, and thus the ratio 
\begin{equation}
r=\frac{\Delta }{\Delta _{total}}=1
\end{equation}
i.e., the variation comes only from the explained sources.

From Eq. (\ref{eq:p1}), if we consider that $y_t=\ln \left\langle
\rho(t+N_{\min})\right\rangle$, $x_t=\ln (t+N_{\min})$, where $N_{\min }$ is
the number of MC steps discarded at the beginning of the simulation (the
first steps). This discard is needed since the universal behavior which we are looking for emerges only
after a time period sufficiently long to avoid the microscopic short-wave behavior \cite{shorttime}. We can define the coefficient of determination as \cite{silva2012} 
\begin{equation}
r=\frac{\sum\limits_{i=N_{\min }}^{N_{MC}}(\overline{\ln \left\langle \rho
\right\rangle }-a-b\ln i)^{2}}{\sum\limits_{i=N_{\min }}^{N_{MC}}(\overline{%
\ln \left\langle \rho \right\rangle }-\ln \left\langle \rho (i)\right\rangle
)^{2}},  \label{eq:coef_det}
\end{equation}
where $N_{MC}$ is the total number of MC steps and $\overline{\ln
\left\langle \rho \right\rangle }=(1/N_{MC})\sum\nolimits_{t=N_{\min
}}^{N_{MC}}\ln \left\langle \rho (t)\right\rangle$. The value of $N_{min}$
depends on the details of the system in study and it is related to the
microscopic time scale, i.e., the time the system needs to reach the
universal behavior in short-time critical dynamics \cite{janssen1989}.

When the system is near the criticality ($y_{c},\rho _{is_{c}}$), we expect
that the order parameter follows a power law behavior which, in $\log \times
\log $ scale, yields a linear behavior and $r$ approaches 1. In this case,
we expect the slope $b$ to be a good estimate of $\beta /\nu _{\parallel }$.
On the other hand, when the system is out of criticality, there is no power
law and $r\simeq 0$. Thus, we are able to use the coefficient of
determination $r$ to look for critical points by considering, for instance,
Eq. (\ref{eq:p1}) for several pairs ($y,\rho _{is}$).

Thus, the idea of the method is very simple: we just need to sweep
the parameter space ($y,\rho_{is}$) and find the points that possess $r
\simeq 1$ and that are, therefore, candidates to continuous phase transition
points. It is important to notice that we can also explore the amplitude of
the method for the study of weak first-order phase transition points \cite
{Zheng2000firstorder,albano2001}. In that case, the transitions exhibit long
correlation lengths and small discontinuities, and therefore, possess a
similar behavior to second-order transitions. So, by following also other
previous works \cite{gennes1975, fernandez1992}, we are able to argue that
the original ZGB model presents a weak first-order transition with two
pseudocritical points, one below and another above its discontinuous phase
transition point. As shown in Refs. \cite{silva2014, dasilva2018,
Zheng2000firstorder, albano2001}, these two points can be determined through
short-time behavior since at these points, the order parameter (and its
higher moments) of the system presents approximate power law behavior.

\section{Results}

\label{sec:results}

The starting point of our study is related to the determination of the
possible critical points present in the ZGB model with inert sites.
Therefore, we first obtain the coefficient of determination $r$ for $10^6$
pairs of $y$ and $\rho_{is}$ through the Eq. (\ref{eq:p1}) by considering $0
\leq y \leq 1$ and $0 \leq \rho_{is} \leq 1$ with $\Delta y = \Delta
\rho_{is} = 10^{-3}$, thus comprising the whole spectrum of possible values
for $y$ and $\rho_{is}$. Hence, we are able to obtain a clue of how is the
behavior of the phase diagram of the model as, for instance, where are the
regions with possible phase transitions and what happens with the continuous
and discontinuous phase transitions of the original model when there exist
inert sites on the lattice.

\begin{figure}[tbh]
\begin{center}
\includegraphics[width=0.95\columnwidth]{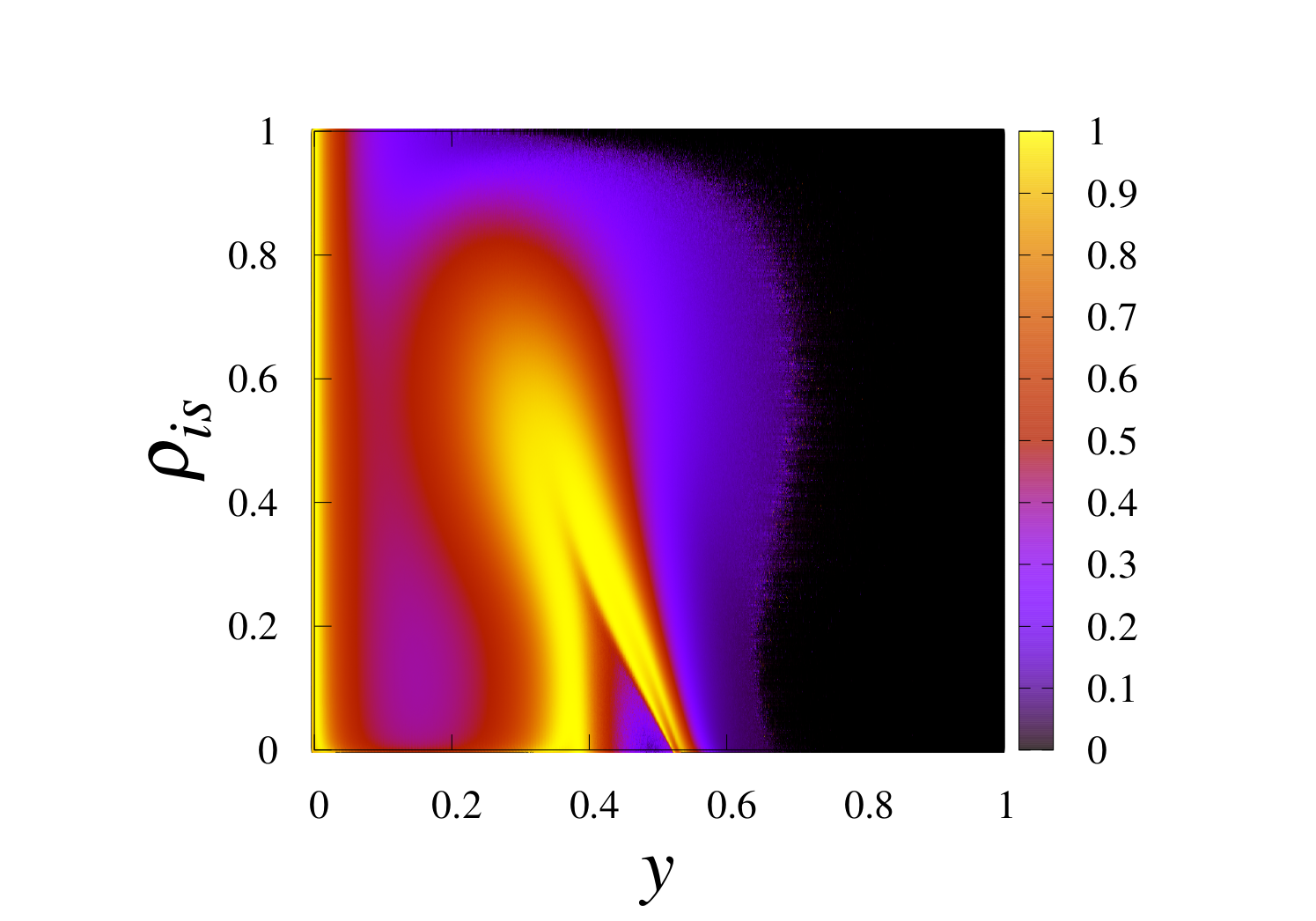}]
\end{center}
\caption{Color map of the coefficient of determination $r$ as function of $y$
and $rho_{is}$ for the density of vacant sites.}
\label{fig:pd_vaz}
\end{figure}

To obtain these diagrams, we consider lattices of linear size $L=80$
, $N_{MC}=500$, $N_{\min}=30$, and $N_{run}=2000$ runs (the number of
samples, i.e., the number of different time evolutions). First of all, it is
important to observe that a massive number of simulations were performed,
and a lattice $L=80$ is the minimum lattice size with both good results and
resoanable machine time. This statement is based on the studies performed
in Ref. \cite{dasilva2018} where we presented a rigorous analysis of the
effects of both lattice size and number of runs on the localization of the
critical points through the coefficient of determination (we invite the
reader to examine Fig. 7 of that reference). We showed that for such
systems, as ZGB model, there are no visual differences for $L$ ranging from
80 to 480 and $N_{run}$ ranging from $10^3$ to $10^4$ runs in the
localization of critical parameters. The number $N_{\min}=30$ was obtained
after the analysis of the time required for the system to reach the
universal behavior. This analysis was performed for the original ZGB model,
i.e., for $\rho_{is}=0$ and y=0.3874. Figure \ref{fig:pd_vaz} shows the
color map of the coefficient of determination $r$ as function of $y$ and $%
\rho _{is}$ for the density of vacant sites $\rho _{V}(t)$ which, in turn,
follows the power law given by Eq. (\ref{eq:p1}). 

As can be seen, the yellow dots are points at which $r$ approaches 1
(therefore, they are candidates to phase transition points) and black dots
are points at which $r$ approaches 0. In addition, we can observe that the
continuous phase transition, that for the original ZGB model, is around $%
y=0.3874$, extends for larger values of $\rho _{is}$ and seems to be
unresponsive to the density of inert sites until $\rho _{is}\simeq 0.3$. For
higher values of $\rho _{is}$, the critical points move to smaller values of 
$y$ until $\rho _{is}\simeq 0.6$. On the other side, the discontinuous phase
transition point, which for the original model is around $y\simeq 0.525$,
seems to be very sensitive to $\rho _{is}$, shifting for decreasing values
of $y$ as $\rho _{is}$ increases. Moreover, for small values of the density
of inert sites, it is possible to observe that the two pseudocritical points
of the original model \cite{fernandes2016, albano2001} are also present, at
least for $\rho _{is}\lesssim 0.2$. Finally, this figure also shows that the
two phase transitions seem to meet each other at the yellow region around $%
\rho _{is}\simeq 0.5$ and that there is no phase transitions when the most
part of the surface is filled with inert sites or when the adsorption rate
of CO molecules is above the discontinuous phase transition point of the
original model. In the first case, the number of inert sites increases and
the adsorption process of O$_{2}$ is hampered by the lack of
nearest-neighbors available to adsorb one of the oxygen atoms. On the other
hand, for higher values of $y$, the system is poisoned with CO molecules.

\begin{figure}[tbh]
\begin{center}
\includegraphics[width=0.95\columnwidth]{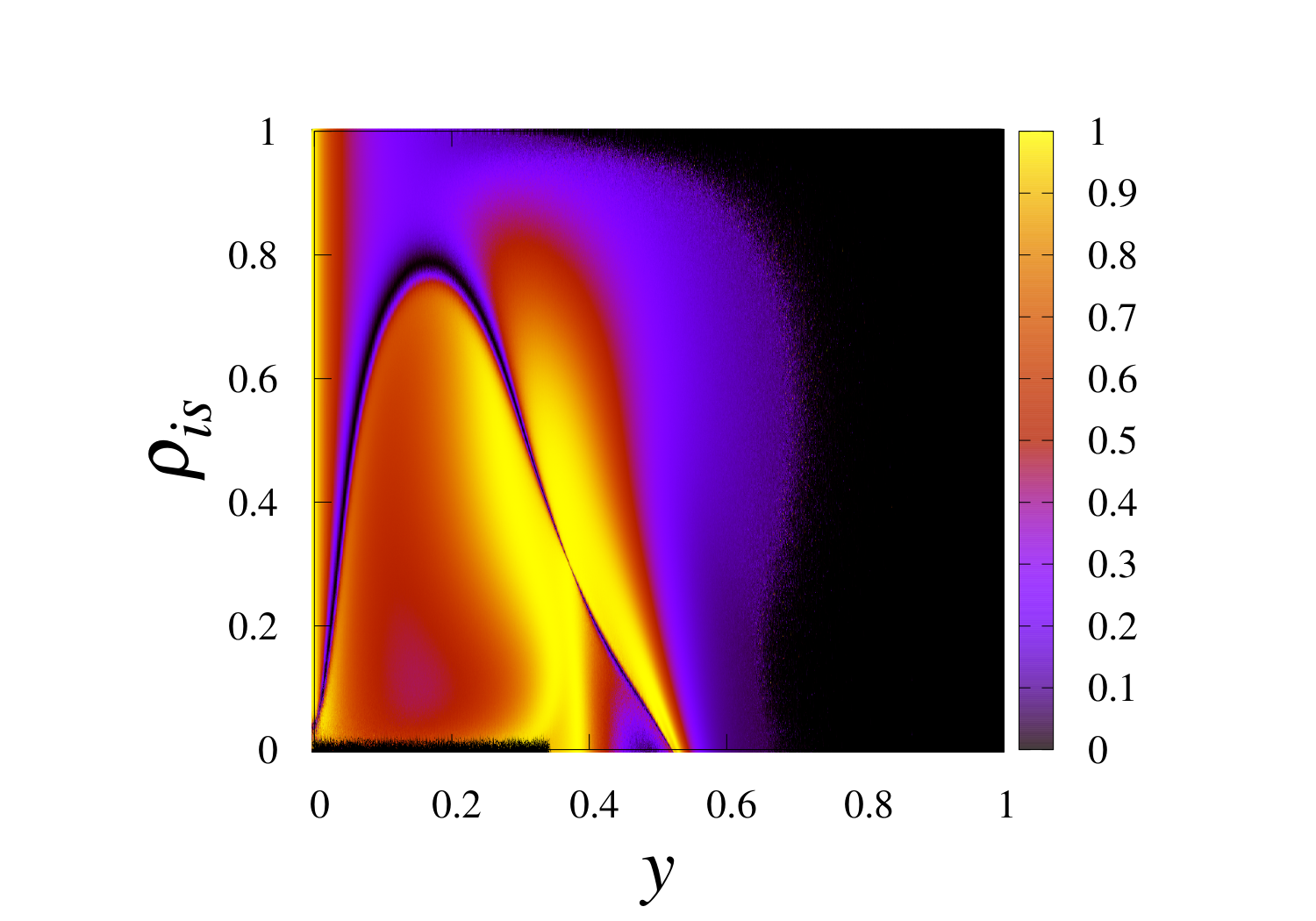}]
\end{center}
\caption{Color map of the coefficient of determination as function of $y$
and $\protect\rho_{is}$ for the density of CO molecules.}
\label{fig:pd_cox}
\end{figure}

Figure \ref{fig:pd_cox} shows the color map for the density of CO molecules (%
$\rho_{\text{CO}}(t)$), which can also be considered as an order parameter
of the model. 

Although the behavior is practically the same presented for $\rho_{\text{V}}(t)$, we are able to gather at least two more important pieces of information to those given above. The first one is that there exists a very thin line separating the points that emerged from the continuous and discontinuous phase transitions of the original model. So, we call Region 1 as the region before this thin line and that therefore comprises the continuous phase transition of the ZGB model, and Region 2 is related to the region after this line comprising the discontinuous phase transition of the original model.

\begin{figure}[tbh]
\begin{center}
\includegraphics[width=0.95\columnwidth]{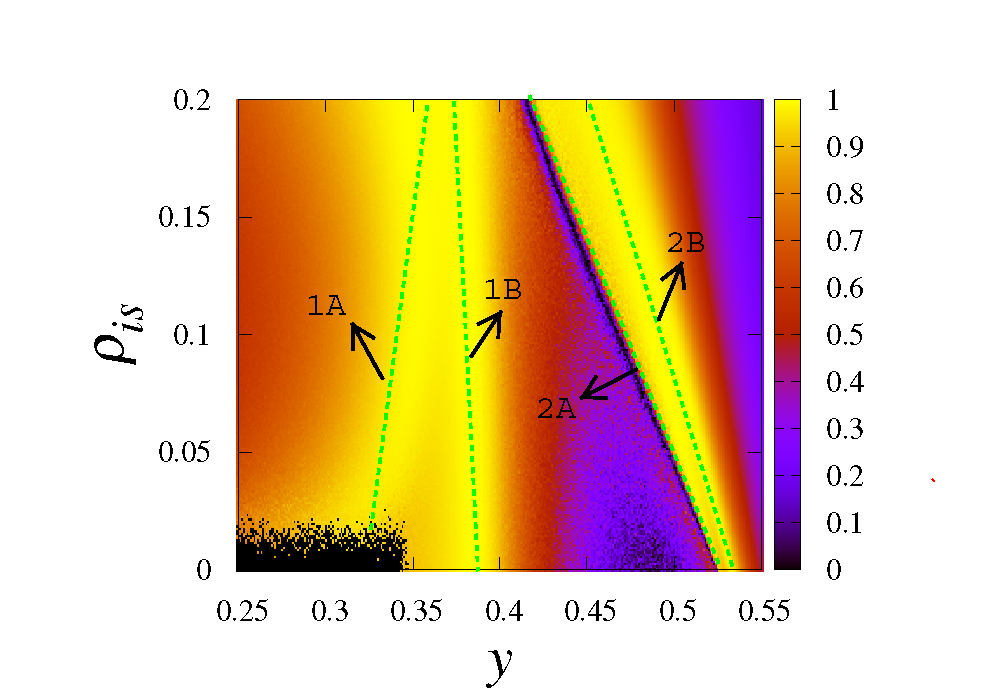}
\end{center}
\caption{Color map of $r$ as function of $y$ and $\protect\rho_{is}$
for $\protect\rho_{\text{CO}}(t)$ and for small values of $\protect\rho_{is}$%
. The green straight lines indicate, approximately, the lines of phase
transition points 1A, 1B, 2A, and 2B. As can be seen, these lines are not
necessarily straight.}
\label{fig:pd_coxzoom}
\end{figure}

The second information is that for small values of $\rho_{is}$, the Region 1 seems to have two lines with $r\simeq 1$, the first line (line 1A) starting at $y\approx 0.3$ (which arises for $\rho_{is}\neq 0$) and the second one (line 1B) starting at $y\approx 0.39$ (around the critical point of the original model). In Addition, Figs. \ref{fig:pd_vaz} and \ref{fig:pd_cox} also show the existence of two lines in Region 2 with $r\simeq 1 $, called lines 2A and 2B. Figure \ref{fig:pd_coxzoom} shows a zoom in both regions for small values of the density of inert sites and these four curves are qualitatively represented by green straight lines.

\begin{figure}[tbh]
\begin{center}
\includegraphics[width=0.8\columnwidth]{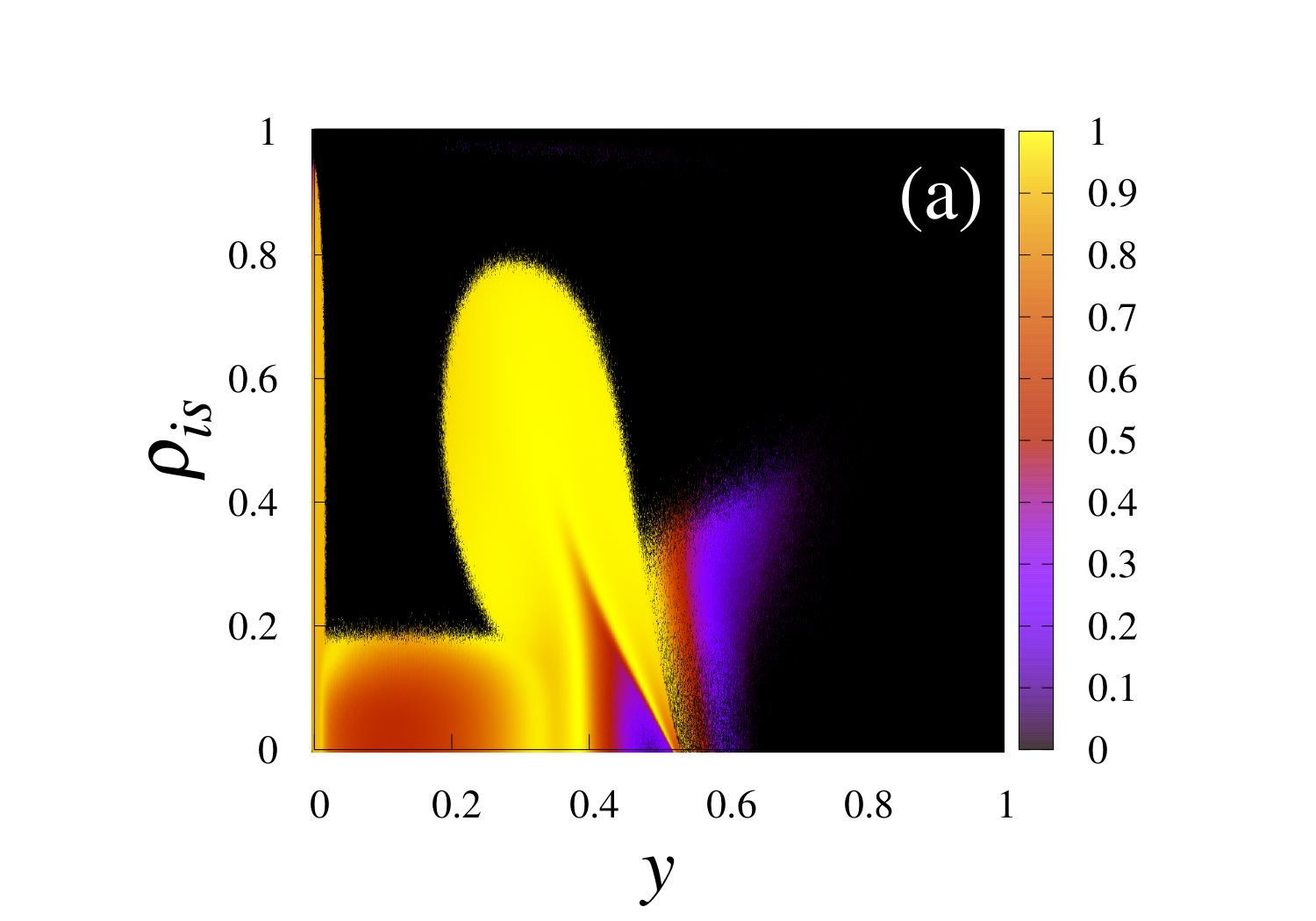} \includegraphics[width=0.8%
\columnwidth]{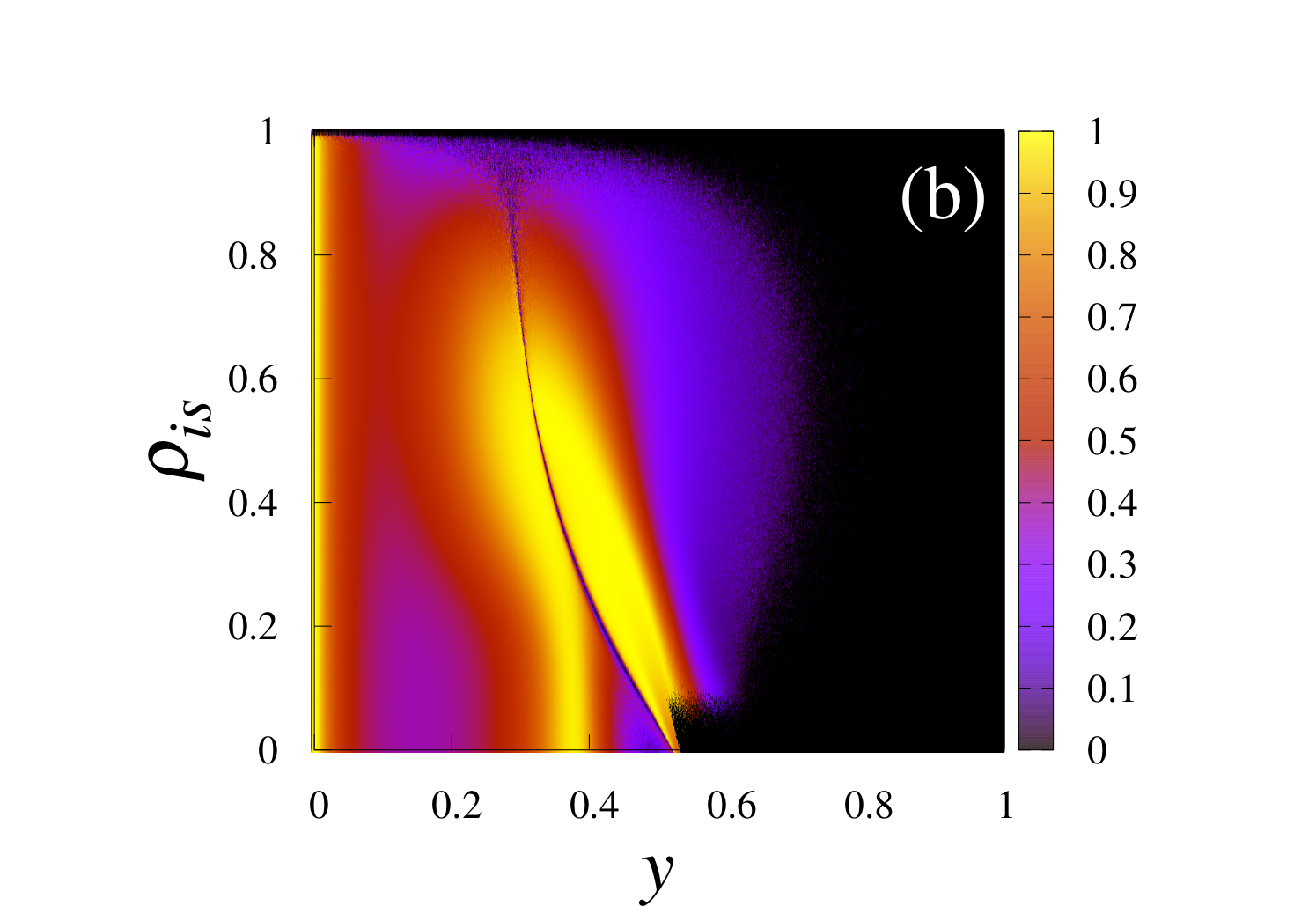}
\end{center}
\caption{Color map of $r$ as function of $y$ and $\protect\rho_{is}$ for (a) 
$\protect\rho_{\text{CO}_2}(t)$ and (b) $\protect\rho_\text{O}(t)$.}
\label{fig:pd_co2_o}
\end{figure}

For completeness, we also obtained the coefficient of determination for the
densities of CO$_2$ molecules, $\rho_{\text{CO}_2}(t)$, and O atoms, $\rho_%
\text{O}(t)$, and their color maps are presented in Fig. \ref{fig:pd_co2_o}
(a) and (b), respectively. As shown, both figures behave similarly to those
presented above for $\rho_\text{V}(t)$ and $\rho_\text{CO}(t)$. 

As can be seen, Figs. \ref{fig:pd_cox} and \ref{fig:pd_co2_o} (b) also show
a region with black points for very small values of $\rho_{is}$. In Fig. \ref%
{fig:pd_cox} the black points are seen when $y\lesssim 0.35$ and in Fig. \ref%
{fig:pd_co2_o} (b) they appear for $y\gtrsim 0.53$. When we deal with the
density of CO molecules, the black points observed in Fig. \ref{fig:pd_cox}
refer to points where the adsorption rates are small and the system goes to
an absorbing state (the available sites are filled with O atoms) at the
beginning of the time evolution. Therefore, $r\simeq 0$ and there is no
power law. As $\rho_{is}$ increases, the system starts to undergo phase
transitions for adsorption rates smaller than that of the original model. If
we look into Fig. \ref{fig:pd_co2_o} (b), we are able to see a similar
behavior at the region of discontinuous phase transition, but in that case,
the system is poisoned by CO molecules at the beginning of the time
evolution and $r\simeq 0$ in that region.

The figures presented above show a very interesting behavior of the ZGB
model when the surface possesses inert sites. The coefficient of
determination was able to capture the regions of possible phase transitions
when both adsorption rate and density of inert sites vary. At this point,
some questions may be raised such as:

\begin{enumerate}
\item Are the yellow dots phase transition points?

\item Are the critical exponents varying with $y$ and $\rho_{is}$?

\item What we could state about the universality class of the model?
\end{enumerate}

To answer these questions, we consider some fixed values of $\rho_{is}$ and
varied the adsorption rate $y$ from 0 to 0.55. In addition, our simulations
are performed for lattices of linear size $L=160$ with 20000 samples and 500
Monte Carlo steps in order to obtain both the coefficient of determination
and critical points with higher precision. Next, we proceed the calculation
of the critical exponents $\beta/\nu_{\parallel}$, $\theta$, $d/z$, and $%
1/\nu_{\parallel}$ from Eqs. (\ref{eq:p1}), (\ref{eq:p2}), (\ref{eq:f2}),
and (\ref{eq:dv}), respectively. With these indexes in hand, we are able to
find the static and dynamic critical exponents separately and compare them
with those values found in literature. It is worth to mention that, from now
on, we take into account the density of CO molecules as the order parameter
of the model. Therefore, only this quantity is used to obtain our main
results presented below by varying the initial conditions of the system.

\begin{figure}[tbh]
\begin{center}
\includegraphics[width=0.95\columnwidth]{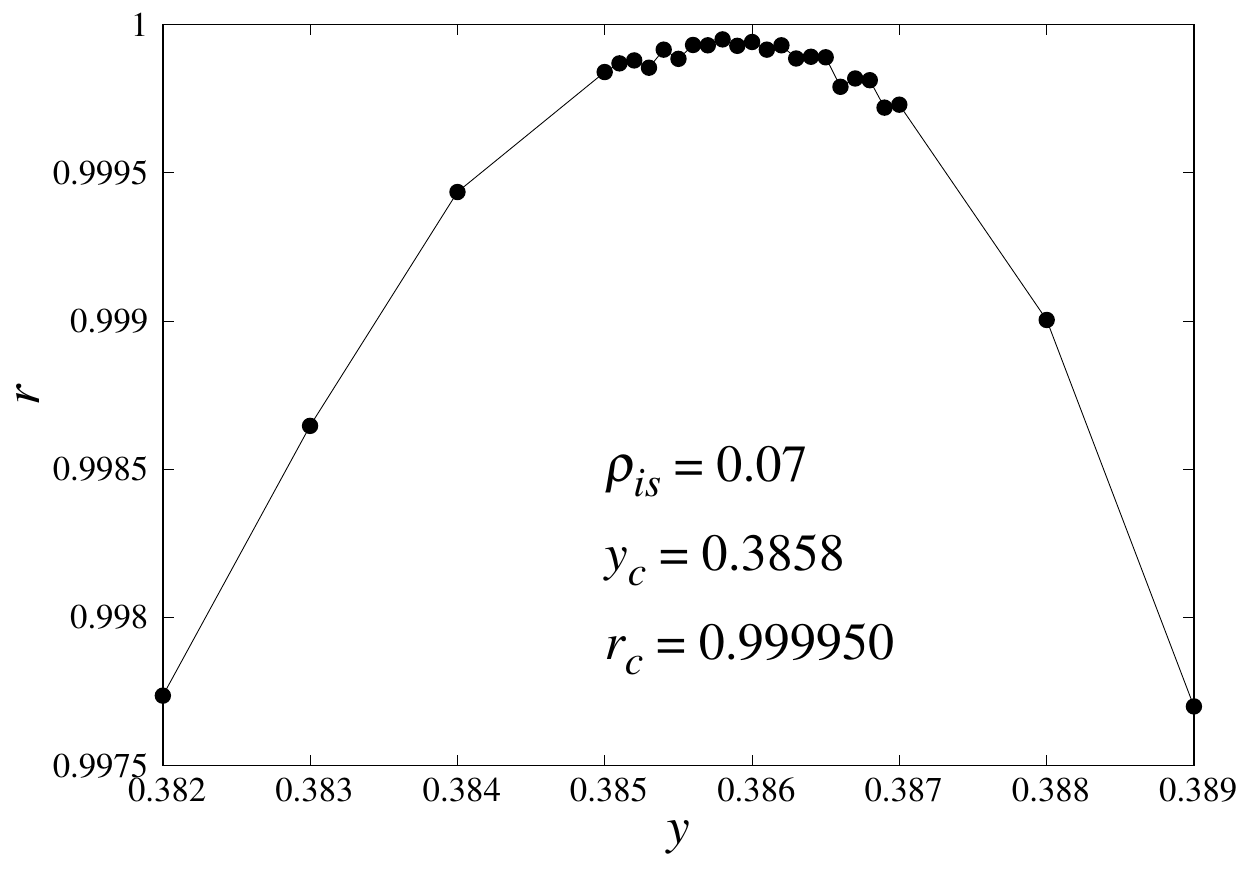}
\end{center}
\caption{Coefficient of determination as function of $y$ for $\protect\rho%
_{is}=0.07$.}
\label{fig:rxy}
\end{figure}

To obtain the coefficient of determination and the critical adsorption
rates, we carry out simulations by following two steps. In the first one, we
obtain each value of $y$ (for the best value of $r$) considering $\Delta
y=10^{-3}$. With these results in hand, we performed new simulations
considering $\Delta y=10^{-4}$ around the former estimate. As an example,
Fig. \ref{fig:rxy} shows the coefficient of determination as function of $y$
for $\rho _{is}=0.07$ for the line 1B. In the first step, we found $%
y_{c}=0.386$ for the higher value of $r$ and, by performing the simulations
with $0.3850\leq y\leq 0.3870$ with $\Delta y=10^{-4}$, we obtained $%
y_{c}=0.3858$ for the critical adsorption rate when $r_{c}=0.999950$. 
This analysis was carried out for all points considered in this work and our
best estimates of the coefficients of determination $r$ and the
corresponding adsorption rates $y$ for several densities of inert sites $%
\rho_{is}$ are shown in Table \ref{tab:tab1}. 
\begin{table}[th]
\caption{Density of inert sites $\protect\rho_{is}$ considered in this work
as well as the adsorption rates $y$ obtained for the best coefficients of
determination $r$. The points are separated into two regions and four lines.}
\label{tab:tab1}%
\begin{tabular}{|c|c|c|c|c|c|c|c|c|}
\hline\hline
\multirow{2}{*}{$\rho_{is}$} & \multicolumn{4}{c|}{Region 1} & 
\multicolumn{4}{c|}{Region 2} \\ \cline{2-9}
& $y_{1A}$ & $r_{1A}$ & $y_{1B}$ & $r_{1B}$ & $y_{2A}$ & $r_{2A}$ & $y_{1B}$
& $r_{2B}$ \\ \hline\hline
0.00 &  &  & 0.3874 & 0.999839 & 0.5271 & 0.999498 & 0.5328 & 0.996256 \\ 
\hline
0.01 &  &  & 0.3871 & 0.999862 & 0.5225 & 0.999731 & 0.5293 & 0.997592 \\ 
\hline
0.02 & 0.3300 & 0.990942 & 0.3868 & 0.999886 & 0.5182 & 0.999810 & 0.5256 & 
0.998490 \\ \hline
0.03 & 0.3389 & 0.992964 & 0.3866 & 0.999898 & 0.5141 & 0.999913 & 0.5221 & 
0.998955 \\ \hline
0.04 & 0.3450 & 0.994137 & 0.3867 & 0.999915 & 0.5107 & 0.999964 & 0.5182 & 
0.999376 \\ \hline
0.05 & 0.3490 & 0.995359 & 0.3865 & 0.999922 & 0.5072 & 0.999976 & 0.5142 & 
0.999591 \\ \hline
0.06 & 0.3520 & 0.996201 & 0.3862 & 0.999941 & 0.5043 & 0.999960 & 0.5102 & 
0.999766 \\ \hline
0.07 & 0.3536 & 0.996961 & 0.3858 & 0.999950 & 0.5012 & 0.999969 & 0.5062 & 
0.999858 \\ \hline
0.08 & 0.3556 & 0.997496 & 0.3854 & 0.999950 & 0.4995 & 0.999958 & 0.4995 & 
0.999958 \\ \hline
0.09 & 0.3575 & 0.998006 & 0.3849 & 0.999958 & 0.4965 & 0.999823 & 0.4965 & 
0.999823 \\ \hline
0.10 & 0.3585 & 0.998261 & 0.3846 & 0.999963 &  &  & 0.4923 & 0.999361 \\ 
\hline
0.15 & 0.3606 & 0.999289 & 0.3812 & 0.999980 &  &  & 0.4745 & 0.995093 \\ 
\hline
0.20 & 0.3553 & 0.999635 & 0.3771 & 0.999993 & 0.4160 & 0.987081 & 0.4563 & 
0.990847 \\ \hline
0.25 & 0.3465 & 0.999794 & 0.3715 & 0.999985 & 0.3976 & 0.998633 & 0.4358 & 
0.988602 \\ \hline
0.30 & 0.3338 & 0.999859 & 0.3670 & 0.999910 & 0.3790 & 0.999354 &  &  \\ 
\hline
0.35 & 0.3209 & 0.999874 &  &  & 0.3710 & 0.997924 &  &  \\ \hline
0.40 & 0.3100 & 0.999859 &  &  & 0.3605 & 0.998146 &  &  \\ \hline
0.50 & 0.2920 & 0.999695 &  &  & 0.3441 & 0.993314 &  &  \\ \hline
0.60 & 0.2770 & 0.998240 &  &  & 0.3337 & 0.965010 &  &  \\ \hline\hline
\end{tabular}%
\end{table}

This table shows, as expected, that the line 1B arises at the critical point
of the original model (for $\rho_{is}=0$), where $y\simeq 0.387$, and the
lines 2A and 2B also arise at the pseudocritical points which are located
close to the discontinuous phase transition point of the ZGB model, $y
\simeq 0.525$. However, the line 1A has a different behavior for small
values of $\rho_{is}$. In fact, as presented in Fig. \ref{fig:pd_coxzoom},
the beginning of this line does not occur for $\rho_{is}=0$ and, therefore,
can not be seen in the original model. Instead, it arises only for $%
\rho_{is} \neq 0$.

Table \ref{tab:tab1} also presents another important result. The adsorption
rates vary according to $\rho _{is}$ for the lines 1A, 2A, and 2B, and
remains stable for the line 1B (at least for small values of $\rho _{is}$),
which is the line starting at the continuous phase transition point of the
original model. Figure \ref{fig:pd_coxzoom} also presents these points as a
vertical line extending until $\rho _{is}\simeq 0.3$. This last behavior had
already been predicted by Hoenicke \textit{et al.} \cite{hoenicke2014}
through the analysis of the behavior of moment ratios of the order parameter.

Another finding is that, the lines 2A and 2B which emerge at the two
pseudocritical points of the standard ZGB model seem to meet each other in a
point for $\rho_{is}\simeq 0.08$ and, from that point on, only one line
probably remains. This result was presented, for the first time, by Hovi 
\textit{et al.} in 1992 \cite{hovi1992}. They showed that, for $\rho_{is}$
greater than 8\%, the first-order phase transition seemed to become
continuous. Hoenicke and Figueiredo \cite{hoenicke2000} had also pointed out
that a continuous phase transition emerged at $\rho_{is}=0.078$, which in
turn, is also in agreement with our result. In another work, Lorenz \textit{%
et al.} \cite{Lorenz2002} showed that the presence of inert sites on the
catalytical surface had the effect of breaking up the surface into regions
of different size producing, at the end, results which corroborate the
appearance of a continuous phase transition even for small values of $%
\rho_{is}$.

In addition, once the inert fraction goes beyond $\rho_{is}\approx
0.5927$ \cite{newman2000}, the percolation threshold for square lattices,
the inert sites will percolate and we are no longer able to observe any
active state. Our results capture this aspect since for $\rho_{is}=0.5$ we
localized a point $y_{2A}=0.3441$ with coefficient of determination $%
r\approx 0.993$ while for $\rho_{is}=0.6$ (after the percolation threshold)
we localized a point $y_{2A}=0.3337$ with worse coefficient of determination 
$r\approx 0.965$. Furthermore, all values found for the coefficient of
determination for $\rho_{is}\leq 0.5$ is greater than 0.99, showing that we
have a sensitive decrease of $r$ when $\rho_{is}$ approaches 0.6.

After these two analyzes, we finish our study of the Region 2 and
turn our attention to the Region 1, which is the region that possesses the
continuous phase transition of the ZGB model (line 1B) and a line of points
which emerge only for higher values of $\rho _{is}$ and around $y=0.3$.
This region possesses several candidates to critical points. However, some
coefficients of determination are not as close to 1 as expected for critical
points, i.e., $r\simeq 1$. So, in the following analysis, we consider only
points with $r \geq 0.9995$ as candidate to phase transition points. Then,
we obtain the critical exponents $\beta/\nu_{\parallel}$, $\theta$, $d/z$,
and $1/\nu_{\parallel}$ from Eqs. (\ref{eq:p1}), (\ref{eq:p2}), (\ref{eq:f2}%
), and (\ref{eq:dv}), respectively. In this study, we also consider lattices
of linear size $L=160$, 20000 samples, and 500 MC steps for Eqs. (\ref{eq:p1}%
), (\ref{eq:p2}), and(\ref{eq:f2}), and 1500 MC steps for Eq. (\ref{eq:dv}).
The error bars are obtained from 5 independent bins.

\begin{figure}[tbh]
\begin{center}
\includegraphics[width=0.95\columnwidth]{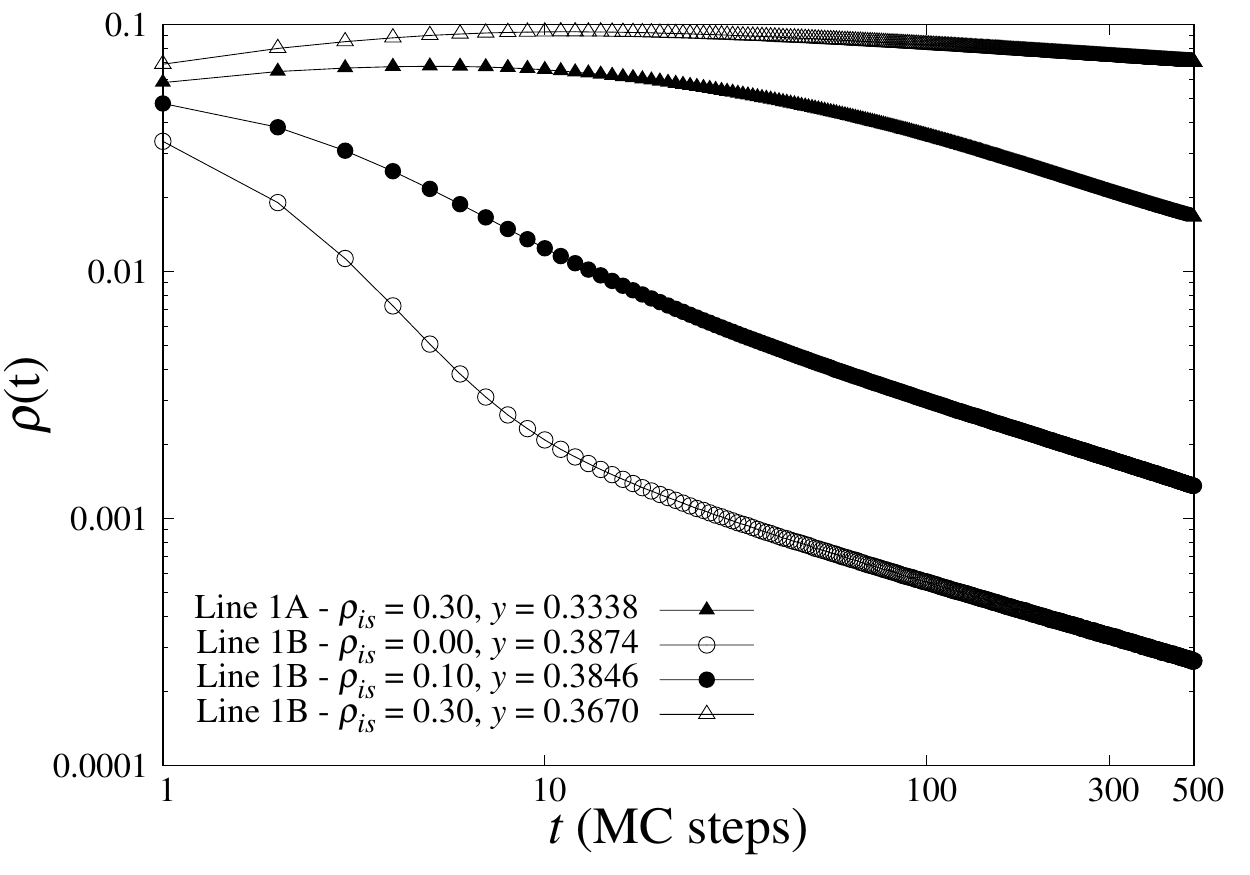}]
\end{center}
\caption{Density of CO molecules as function of $t$ for four critical points
of the model. One curve is obtained from a critical point belonging to the
line 1A and three curves are taken from the line 1B. All curves present
linear behavior for $t>100$ MC steps. These curves represent one bin from
the five ones considered in this work.}
\label{fig:rhodecay}
\end{figure}

Figure \ref{fig:rhodecay} shows the power law decay of the density of CO
molecules as function of $t$ (Eq. (\ref{eq:p1})) in $\log \times \log $
scale for four points: one point from line 1A and three points taken from
line 1B. 
As can be seen, after 100 MC steps, all curves follow linear behaviors
meaning that, according to the short-time dynamics, these points are in fact
critical points. From the slope of these curves, we are able to obtain the
critical exponent $\beta/\nu_{\parallel}$.

By following this procedure for the other power law equations presented
above, we calculate the corresponding critical exponents after discarding
some initial MC steps. For the Eq. (\ref{eq:p1}), we discarded the first 100
MC steps and for Eqs. (\ref{eq:p2}) and (\ref{eq:f2}) we discarded the first
200 MC steps. To obtain the linear behavior of the Eq. (\ref{eq:dv}), we
needed 1500 MC steps and the first 700 were discarded.

Table \ref{tab:tab2} presents all the critical exponents estimated in this
work with the respective error bars. 
\begin{table}[th]
\caption{Static and dynamic critical exponents obtained for several critical
points of the ZGB model with inert sites.}%
\begin{tabular}{|c|c|c|c|c|c|c|c|c|}
\hline\hline
\multirow{3}{*}{$\rho_{is}$} & \multicolumn{8}{c|}{Region 1} \\ \cline{2-9}
& \multicolumn{4}{c|}{Line 1A} & \multicolumn{4}{c|}{Line 1B} \\ \hline\hline
& $\beta $ & $\nu _{\parallel }$ & $z$ & $\theta $ & $\beta $ & $\nu
_{\parallel }$ & $z$ & $\theta $ \\ \hline
0,00 &  &  &  &  & 0.6162(90) & 1.360(17) & 1.771(22) & 0.223(15) \\ \hline
0.01 &  &  &  &  & 0.642(19) & 1.392(37) & 1.744(16) & 0.224(11) \\ \hline
0.02 &  &  &  &  & 0.635(10) & 1.344(20) & 1.727(30) & 0.213(20) \\ \hline
0.03 &  &  &  &  & 0.639(19) & 1.340(36) & 1.736(32) & 0.198(21) \\ \hline
0.04 &  &  &  &  & 0.640(19) & 1.367(37) & 1.732(15) & 0.2183(91) \\ \hline
0.05 &  &  &  &  & 0.633(17) & 1.347(35) & 1.738(41) & 0.211(27) \\ \hline
0.06 &  &  &  &  & 0.642(26) & 1.352(51) & 1.747(36) & 0.194(22) \\ \hline
0.07 &  &  &  &  & 0.655(15) & 1.355(27) & 1.744(20) & 0.180(14) \\ \hline
0.08 &  &  &  &  & 0.661(12) & 1.352(22) & 1.747(23) & 0.160(10) \\ \hline
0.09 &  &  &  &  & 0.668(15) & 1.346(27) & 1.753(34) & 0.148(22) \\ \hline
0.10 &  &  &  &  & 0.682(28) & 1.383(54) & 1.741(34) & 0.162(22) \\ \hline
0.15 &  &  &  &  & 0.748(25) & 1.512(48) & 1.794(26) & 0.126(15) \\ \hline
0.20 &  &  &  &  & 0.733(36) & 1.815(85) & 1.934(68) & 0.227(34) \\ \hline
0.25 & 1.664(81) & 2.60(12) & 6.4(1.5) & -0.971(73) & 0.651(35) & 2.43(13) & 
2.176(50) & 0.383(21) \\ \hline
0.30 & 1.459(72) & 3.04(15) & 6.9(1.7) & -0.669(72) &  &  &  &  \\ \hline
0.35 & 1.135(31) & 3.262(85) & 9.1(3.3) & -0.307(47) &  &  &  &  \\ 
\hline\hline
\end{tabular}%
\label{tab:tab2}
\end{table}

This table shows only three points for the line 1A which in turn present
critical exponents with huge error bars. Although a power law behavior is
found for Eq. (\ref{eq:p1}) which lead to high values of $r$, the other
equations do not follow the behavior expected for critical points (linear
behavior in $\log \times \log $ scale), presenting huge fluctuations. As can
be seen, the exponents $\beta $, $\nu _{\parallel }$, and $z$ have very high
values, and the exponent $\theta $ is negative. So, the behavior of the
functions, the huge fluctuations, and the values obtained for the exponents
in this work prevented us to assert that the line 1A is a line of critical
points leaving this subject as an open question which should be addressed in
a future work. As the coefficient of determination is a refinement method
which has proven to be very efficient in determining critical points of
several models, including the ZGB model, we believe that the $r-$values
obtained mainly for the three points of the line 1A presented in Table \ref%
{tab:tab2} are at least a clue that there is something in this region that
other approaches have not been able to observe.

On the other hand, the critical exponents obtained for the line 1B are much
more consistent. We can observe that the exponents for $\rho_{is}=0.00$ are
very close to the values obtained for the standard ZGB model. For instance,
by using both epidemic and poisoning-time analyzes, Voigt and Ziff \cite%
{voigt1997} obtained $\beta=0.584(4)$, $\nu_{\parallel}=1.295(6)$, $z=1.76(3)
$, and $\theta=0.2295(10)$, and recently, Fernandes \textit{et al.} obtained 
$\beta=0.586(7)$, $\nu_{\parallel}=1.292(15)$, $z=1.756(3)$, and $%
\theta=0.231(3)$, by means of short-time Monte Carlo simulations.

For small values of $\rho _{is}$, our results for the static critical
exponent $\beta $ show a tendency of increasing and this finding is
supported by the results of Hoenicke \textit{et al.} \cite{hoenicke2014}.
For instance, our estimates of $\beta $ for $\rho _{is}=0.01$ and $\rho
_{is}=0.10$ are $\beta =0.642(19)$ and $\beta =0.682(28)$, respectively,
while in Ref. \cite{hoenicke2014}, the authors found $\beta =0.623(7)$ and $%
\beta =0.707(7)$, respectively, which are in agreement with each other. It
is interesting to observe that the static critical exponent $\nu _{\parallel
}$ and the dynamic critical exponent $\theta $ seem to vary only for $\rho
_{is}\geq 0.15$.

\section{Conclusions}

\label{sec:conclusions}

In this work, we carried out nonequilibrium Monte Carlo simulations along
with a refinement method in order to obtain the coefficient of determination
of the ZGB model with inert sites. We presented diagrams of the possible
phase transition points of the model through color maps of the coefficient
of determination as function of the adsorption rates of carbon monoxide, $y$%
, and of the densities of inert sites, $\rho_{is}$. These diagrams showed
two regions of phase transitions: one continuous and another corresponding
to an extension of the discontinuous point existing for $\rho_{is}=0$.
The results showed that the continuous phase transition of the
original model is weakly influenced by the inclusion of inert sites on the
catalytic surface. With the diagrams in hand, we turned our attention to
the refinement of the phase transition points by fixing some values of
density of inert sites and using larger lattices and samples in order to
obtain $y$ with higher precision. Finally, we focused our attention to the
region of continuous phase transitions and calculated the static critical
exponents $\beta$, $\nu_{\parallel}$ and $\nu_{\perp }$ and the dynamic ones 
$z$ and $\theta $ for some specific points.

\section*{Acknowledgments}

R. da Silva thanks CNPq for financial support (grant 310017/2015-7). This
research was partially carried out using the computational resources of the
Center for Mathematical Sciences Applied to Industry (CeMEAI) funded by
FAPESP (grant 2013/07375-0).

\end{document}